\documentclass[conference]{IEEEtran}
\IEEEoverridecommandlockouts

\usepackage{cite}
\usepackage{amsmath,amssymb,amsfonts}
\usepackage{algorithmic}
\usepackage{graphicx}
\usepackage{physics}
\usepackage{comment}
\usepackage{textcomp}
\usepackage{xcolor}
\usepackage{booktabs}
\usepackage{subcaption}
\usepackage{hyperref}

\usepackage{algorithm}
\usepackage{listings}
\usepackage[capitalize]{cleveref}
\crefname{equation}{Eq.}{Eqs.}
\Crefname{equation}{Eq.}{Eqs.}

\crefname{subfigure}{Fig.}{Figs.}
\Crefname{subfigure}{Fig.}{Figs.}

\def\BibTeX{{\rm B\kern-.05em{\sc i\kern-.025em b}\kern-.08em
    T\kern-.1667em\lower.7ex\hbox{E}\kern-.125emX}}

\lstdefinestyle{dimacs}{
    basicstyle=\ttfamily\small,
    frame=single,
    framesep=4pt,
    numbers=left,
    numberstyle=\scriptsize\color{gray},
    numbersep=8pt,
    xleftmargin=1.5em,
    breaklines=true,
    columns=fullflexible,
    keepspaces=true,
    showstringspaces=false,
    aboveskip=6pt,
    belowskip=6pt,
    morekeywords={cnf, qcnf}
    backgroundcolor=\color{gray!5},
    rulecolor=\color{black!40},
    commentstyle=\color{green!50!black},
    morecomment=[l]{c\ }
}

\begin{document}

\title{
SAQC: A SAT-Aware Compilation Framework for QAOA-Based Quantum Optimization
}
\author{
Till~Schnittka\textsuperscript{$\Delta$},
Julie~Maria~Raju\textsuperscript{$\Cup$},
Abhoy~Kole\textsuperscript{$\Delta$},
Rolf~Drechsler\textsuperscript{$\Delta$,$\Cup$}\\
 
 \textsuperscript{$\Delta$}\textit{Cyber-Physical Systems, DFKI, Bremen, Germany} \\ 
 \textsuperscript{$\Cup$}\textit{Institute of Computer Science, University of Bremen, Bremen, Germany}\\  
 \{abhoy.kole, majd.assaad, till.schnittka\}@dfki.de, drechsler@uni-bremen.de }

\title{SAQC: A SAT-Aware Compilation Framework for QAOA-Based Quantum Optimization
\\

}

\author{
\IEEEauthorblockN{Till~Schnittka}
\IEEEauthorblockA{\textit{Cyber-Physical Systems} \\
\textit{DFKI GmbH}\\
Bremen, Germany \\
till.schnittka@dfki.de}
\and
\IEEEauthorblockN{Julie~Maria~Raju}
\IEEEauthorblockA{\textit{Institute of Computer Science} \\
\textit{University of Bremen}\\
Bremen, Germany \\
jraju@uni-bremen.de}
\and
\IEEEauthorblockN{Abhoy~Kole}
\IEEEauthorblockA{\textit{Cyber-Physical Systems} \\
\textit{DFKI GmbH}\\
Bremen, Germany \\
abhoy.kole@dfki.de}
\and
\IEEEauthorblockN{Rolf~Drechsler}
\IEEEauthorblockA{\textit{Institute of Computer Science} \\
\textit{University of Bremen / DFKI GmbH}\\
Bremen, Germany \\
drechsler@uni-bremen.de}
}


\maketitle

\begin{abstract}
The Quantum Approximate Optimization Algorithm (QAOA) is a promising variational approach for solving Boolean satisfiability (SAT) problems. Existing SAT-to-QAOA workflows first translate Boolean formulas into penalty Hamiltonians before applying generic quantum compilation, thereby discarding SAT-specific information such as clause structure, logical operators, and variable dependencies. Consequently, subsequent compiler optimizations cannot exploit the underlying Boolean formulation. This paper presents SAQC, a SAT-aware compilation framework that introduces a SAT Intermediate Representation (SIR) to preserve Boolean structure during compilation. The proposed framework supports both CNF and XOR-extended CNF (eCNF) formulations and performs SAT-aware optimizations including clause rewriting, dependency analysis, and clause scheduling prior to quantum lowering. Building on the optimized SIR, SAQC provides a unified framework for both penalty Hamiltonian generation and direct clause-to-ansatz synthesis, together with ansatz optimizations based on relative-phase decomposition, ancilla-assisted synthesis, and dynamic uncomputation. Experimental results demonstrate significant reductions in circuit depth, two-qubit gate count, and compilation time compared with conventional Hamiltonian-based workflows while remaining compatible with existing quantum compilation frameworks.
\end{abstract}

\begin{IEEEkeywords}
Quantum Computing, QAOA, Quantum Compilation, Boolean Satisfiability (SAT), Intermediate Representation
\end{IEEEkeywords}

\section{Introduction}

Boolean satisfiability (SAT) is one of the most fundamental NP-complete problems~\cite{10.1145/800157.805047} and forms the computational backbone of numerous Electronic Design Automation (EDA) tasks, including formal verification, equivalence checking, automatic test generation, logic synthesis, and optimization~\cite{Biere2009HandbookSAT,drechsler2015formal}. While modern classical SAT solvers have achieved remarkable performance on many practical instances~\cite{10.1007/978-3-540-24605-3_37,Biere2020CAD}, the exponential worst-case complexity of SAT continues to motivate the exploration of alternative computational paradigms~\cite{farhi2014quantumapproximateoptimizationalgorithm,10.1145/3729229}.

Among emerging quantum approaches, the Quantum Approximate Optimization Algorithm (QAOA)~\cite{farhi2014quantumapproximateoptimizationalgorithm} has attracted considerable attention as a promising variational algorithm for solving combinatorial optimization problems on noisy intermediate-scale quantum (NISQ) devices~\cite{Preskill2018quantumcomputingin}. Recent works have demonstrated QAOA formulations for SAT and more general constraint satisfaction problems by mapping Boolean formulas into cost Hamiltonians whose ground states correspond to satisfying assignments~\cite{PRXQuantum.5.030348,a12020034,10.3389/fphy.2014.00005}. These studies primarily focus on Hamiltonian construction, parameter optimization, mixer design, or theoretical performance analysis~\cite{PhysRevX.10.021067,crooks2018performancequantumapproximateoptimization}.
\begin{figure}[!t]
    \centering
    \includegraphics[width=0.95\columnwidth]{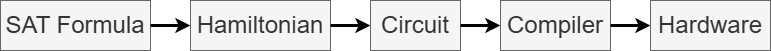}
    \caption{Existing SAT-QAOA Compilation Flow.}
    \label{fig:SAT-Compilation}
\end{figure}
However, comparatively little attention has been devoted to the compilation of SAT-derived QAOA circuits. Existing SAT-QAOA workflows (see~\autoref{fig:SAT-Compilation}) translate Boolean formulas into cost Hamiltonians, from which quantum circuits are synthesized and optimized using generic quantum compilers such as Qiskit~\cite{javadiabhari2024quantumcomputingqiskit}. Although mathematically equivalent, the Hamiltonian representation no longer explicitly preserves the Boolean semantics of the original SAT instance. Consequently, compiler optimizations operate solely at the quantum gate level and cannot exploit SAT-specific structural information.

To address this limitation, we propose SAQC, a SAT-aware compilation framework that introduces a SAT Intermediate Representation (SIR) to preserve Boolean semantics prior to Hamiltonian generation. Following the compiler design philosophy exemplified by intermediate representations such as MLIR~\cite{9370308}, the proposed SIR enables SAT-aware analyses and transformations before lowering Boolean formulas to Hamiltonians, thereby exposing optimization opportunities that are unavailable in conventional SAT-to-QAOA workflows.
Specifically, SAQC provides:
\begin{enumerate}
    \item a unified SIR for both CNF and XOR-extended CNF (eCNF) formulations;
    \item SAT-aware analysis, including clause rewriting and dependency analysis;
    \item optimized Hamiltonian generation and gate synthesis;
    \item SAT-aware scheduling for generating depth-efficient QAOA circuits; and
    \item a comprehensive experimental evaluation assessing the scalability of the proposed framework and the quality of the generated Hamiltonians and QAOA ansätze.
\end{enumerate}


The remainder of this paper is organized as follows. Section II provides the necessary background. Section III presents the proposed SAQC framework. Section IV reports the experimental evaluation. Finally, Section V concludes the paper.
\section{Background}
\subsection{Quantum Approximate Optimization Algorithm}
The Quantum Approximate Optimization Algorithm (QAOA)~\cite{farhi2014quantumapproximateoptimizationalgorithm} is a hybrid quantum--classical variational algorithm for solving combinatorial optimization problems on noisy intermediate-scale quantum (NISQ) devices~\cite{Preskill2018quantumcomputingin}. Given a problem defined over $n$ binary variables, the objective function is encoded as a diagonal cost Hamiltonian
\begin{equation}
H_C=\sum_{x\in\{0,1\}^{n}}H(x)\ket{x}\bra{x},
\label{eq:cost_hamiltonian}
\end{equation}
whose ground state corresponds to the optimal solution of the classical optimization problem.

QAOA initializes the quantum state in the uniform superposition

\begin{equation}
\ket{+}^{\otimes n} = \frac{1}{\sqrt{2^n}} \sum_{x\in\{0,1\}^{n}}\ket{x},
\label{eq:initial_state}
\end{equation}

and alternates $p$ rounds of problem-dependent and mixer evolutions. The mixer Hamiltonian~\cite{a12020034} is commonly chosen as
\begin{equation}
H_B=\sum_{j=1}^{n}X_j,
\label{eq:mixer_hamiltonian}
\end{equation}
where $X_j$ denotes the Pauli-$X$ operator acting on qubit $j$. The resulting variational state is
\begin{equation}
\ket{\psi_p(\boldsymbol{\gamma},\boldsymbol{\beta})} = \prod_{k=0}^{p-1} e^{-i\beta_k H_B} e^{-i\gamma_k H_C}\ket{+}^{\otimes n},
\label{eq:qaoa_state}
\end{equation}
where $\boldsymbol{\gamma},\boldsymbol{\beta}\in\mathbb{R}^p$ are optimized by a classical optimizer to minimize the expectation value
\begin{equation}
E_p(\boldsymbol{\gamma},\boldsymbol{\beta})=\bra{\psi_p}H_C\ket{\psi_p}.
\label{eq:qaoa_objective}
\end{equation}
After optimization, repeated measurements of $\ket{\psi_p(\boldsymbol{\gamma}^{*},\boldsymbol{\beta}^{*})}$ produce candidate solutions with high probability of approximating the optimum.

From a compilation perspective, both the cost Hamiltonian and the Boolean structure from which it is derived influence the resulting QAOA circuit~\cite{PRXQuantum.5.030348}. Different encodings of an equivalent Boolean problem can produce Hamiltonians with different locality, Pauli-term counts, and commutation structures, leading to significant variations in circuit depth, two-qubit gate count, routing overhead, and hardware efficiency.Consequently, QAOA efficiency depends not only on variational parameter optimization and Hamiltonian encoding, but also on how Boolean constraints are lowered into the cost unitary.

\subsection{Boolean Satisfiability(SAT)}\label{sec:SAT}

The Boolean satisfiability (SAT) problem determines whether there exists an assignment to a set of Boolean variables $\mathbf{x}=\{x_1,\ldots,x_n\}$ that satisfies a Boolean formula. In its standard representation, the formula is expressed in conjunctive normal form (CNF)~\cite{Biere2009HandbookSAT}, meaning it is a conjunction of $m$ clauses,
\begin{equation}
\Phi(\mathbf{x}) = \bigwedge_{j=1}^{m} C_j,\label{eq:cnf}
\end{equation}
where each clause $C_j$ is a disjunction of $k$ literals. A literal is either a variable $x_i$ or its negation $\neg x_i$. A clause is satisfied if at least one of its literals evaluates to true, while the full formula is satisfied if and only if all clauses are satisfied.

To solve SAT using QAOA, the Boolean formula is mapped to a cost Hamiltonian~\cite{10.3389/fphy.2014.00005,PRXQuantum.5.030348} by assigning each clause a penalty Hamiltonian that evaluates to zero for satisfying assignments and incurs a positive energy penalty otherwise. The resulting cost Hamiltonian is
\begin{equation}\label{eq:Hamiltonian}
H_C=\sum_{j=1}^{m}H_j,
\end{equation}
where $H_j$ denotes the penalty Hamiltonian corresponding to clause $C_j$. For example, consider the clause 
$C_j=(x_1\vee x_2\vee x_3)$, that is violated only when all three variables evaluate to false. The corresponding penalty Hamiltonian can therefore be expressed as
\begin{equation}
H_j = \prod_{x_i \in C_j} \frac{1 + Z_i}{2},
\end{equation}
where $Z_i$ is the Pauli-$Z$ operator acting on qubit $i$. For a negated literal $\neg x_i$, the corresponding projector becomes
$\frac{I-Z_i}{2}$.

Besides CNF, Boolean formulas may also be represented using XOR-extended CNF (eCNF)~\cite{Biere2009HandbookSAT,10.1145/3729229}, in which XOR constraints are represented explicitly rather than expanded into multiple CNF clauses. Such formulations preserve the underlying Boolean structure and often provide a more compact representation for parity-constrained and cryptographic SAT instances.

\subsection{Related Work and Motivation}

QAOA has been widely investigated for SAT and related constraint satisfaction problems. Early work established the QAOA framework for combinatorial optimization~\cite{farhi2014quantumapproximateoptimizationalgorithm}, while subsequent studies developed SAT-specific~\cite{PRXQuantum.5.030348} and general CSP formulations~\cite{boulebnane2024applyingquantumapproximateoptimization}. These works primarily focus on Hamiltonian construction, variational optimization, mixer design, and algorithmic performance.

Hamiltonian formulations for Boolean functions have also been extensively studied, including Ising encodings of NP-complete problems~\cite{10.3389/fphy.2014.00005} and general mappings for constrained optimization~\cite{10.1145/3478519}, providing the foundation for Hamiltonian-based quantum optimization.

Compiler support for QAOA remains comparatively limited. Existing approaches, such as 2QAN~\cite{lao20212qanquantumcompiler2local}, optimize Hamiltonian-derived circuits, while general-purpose frameworks including Qiskit~\cite{javadiabhari2024quantumcomputingqiskit}, t$\ket{\text{ket}}$~\cite{8f358dfb43c441e1976b7a04bd86dc1b}, and PennyLane~\cite{bergholm2022pennylaneautomaticdifferentiationhybrid} perform gate-level synthesis and hardware-aware transpilation.

In contrast, SAQC performs compilation before Hamiltonian synthesis. By introducing a SAT Intermediate Representation (SIR), it preserves the Boolean structure of both CNF and XOR-extended CNF (eCNF) formulations, enabling SAT-aware analyses, optimizations, and unified generation of penalty Hamiltonians and QAOA ansätze.

\section{Proposed SAT-Aware Compilation Framework}
\begin{figure}[!t]
    \centering
    \includegraphics[width=0.92\columnwidth]{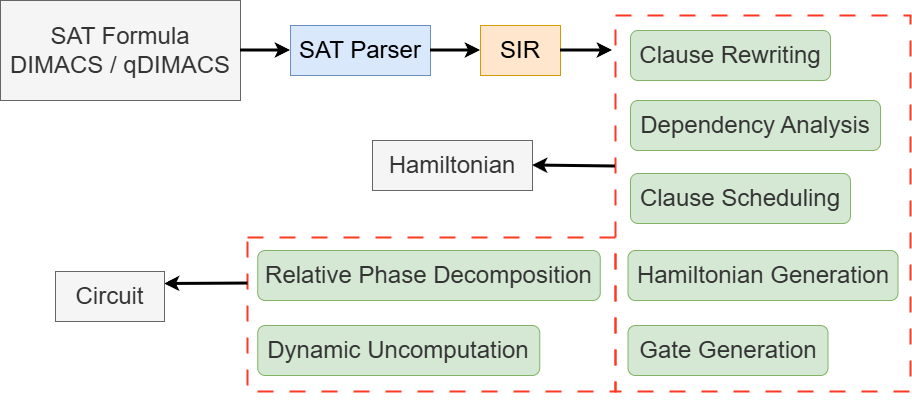}
    \caption{Overview of the SIR-based compilation flow.}
    \label{fig:SAT-prop-Compilation}
\end{figure}

\autoref{fig:SAT-prop-Compilation} presents the overall workflow of the proposed SAT-Aware Quantum Compilation (SAQC) framework. Starting from a SAT instance represented in DIMACS or qDIMACS format, SAQC first constructs a SAT Intermediate Representation (SIR), which preserves the logical structure of the Boolean formula throughout compilation. The SIR forms the basis for compiler analyses and optimization passes, including clause rewriting, dependency analysis, and clause scheduling. The optimized representation is subsequently lowered through dedicated CNF- and eCNF-specific translation rules to generate both the corresponding penalty Hamiltonian and the parameterized QAOA ansatz. Finally, the generated ansatz is optimized using relative-phase gate transformations, ancilla-assisted decompositions, and dynamic uncomputation before hardware execution.

\subsection{A SAT Intermediate Representation}

The proposed SAQC framework accepts Boolean formulas represented either in the standard DIMACS format for CNF or in the extended qDIMACS format supporting XOR constraints (eCNF). Regardless of the input representation, the first stage of SAQC constructs a SIR for further processing.

A conventional DIMACS file consists of a sequence of CNF clauses, where each clause is represented as a list of signed integer literals terminated by~0. Positive and negative integers denote positive and negated literals, respectively. An example is shown in \autoref{lst:dimacs}.
\noindent
\begin{minipage}{0.98\columnwidth}
\begin{lstlisting}[style=dimacs, numbers=none, caption={Example DIMACS (CNF) representation.},label={lst:dimacs}]
p cnf 10 15
c A -> 9
c B -> 10
...
-4 -5 2 0
-9 -10 -5 3 0
...
\end{lstlisting}
\end{minipage}

In contrast, qDIMACS extends the DIMACS syntax by explicitly representing XOR constraints without expanding them into multiple CNF clauses. The separator ``@'' denotes XOR operands, while ``!'' represents the Boolean constant~1. An example is shown in \autoref{lst:qdimacs}.
\noindent
\begin{minipage}{0.98\columnwidth}
\begin{lstlisting}[style=dimacs, numbers=none, caption={Example qDIMACS (eCNF) representation.},label={lst:qdimacs}]
p qcnf 8 5
c A -> 9
c B -> 10
...
! @ 3 @ 6 7 8 0
! @ 4 @ -6 -7 0
...
\end{lstlisting}
\end{minipage}

Instead of processing these formats independently, SAQC translates both representations into a common SIR. As summarized in Table~\ref{tab:sir}, the SIR represents each clause as a logical operator together with its associated literals or subexpressions, providing a unified representation for both CNF and eCNF formulations. This abstraction enables subsequent optimization and lowering passes to operate independently of the original input syntax.
{
\setlength{\tabcolsep}{4pt} 
\begin{table}[htbp!]
\centering
\caption{Logical constructs represented in the SIR.}
\label{tab:sir}
\begin{tabular}{llll}
\toprule
Construct & Example & Construct & Example\\
\midrule
Positive literal & $x_i$ & CNF clause & $x_1\vee x_2\vee \neg x_3$\\
Negated literal & $\neg x_i$ & XOR clause & $x_1\oplus x_2\oplus x_3$\\
Boolean constant & $1$ & XOR-of-products & $x_1\oplus(x_2\land x_3)$\\
\bottomrule
\end{tabular}
\end{table}}

Representing both CNF and eCNF clauses using a common intermediate representation enables subsequent optimization, including clause rewriting, dependency analysis, clause scheduling, Hamiltonian generation, and direct clause-to-ansatz synthesis.

\subsection{SAT-aware Optimization Passes}

After constructing the SIR, SAQC performs three optimization passes before lowering the Boolean formula to a penalty Hamiltonian or a QAOA ansatz: (i) clause rewriting, (ii) dependency analysis, and (iii) clause scheduling.

\subsubsection{Clause Rewriting}

The complexity of both the penalty Hamiltonian and the generated ansatz depends on the size of the Boolean clauses. Large conjunctions in eCNF clauses produce high-order Pauli interactions and large multi-controlled Toffoli gates, both of which are expensive to synthesize. To alleviate this, SAQC introduces auxiliary variables to recursively decompose large clauses into smaller subexpressions. Since each Boolean variable corresponds to one qubit, the additional variables naturally translate into auxiliary qubits.

For an eCNF clause, the rewriting rule is as follows (assuming $c$ is an unused variable):

\begin{align}
a \oplus (b_0 \land \cdots \land b_n)
&\rightarrow
(a \oplus (b_0 \land \cdots \land b_{\lfloor n/2\rfloor}\land c))
\nonumber\\
&
\land
(\overline c \oplus (b_{\lfloor n/2\rfloor+1}\land\cdots\land b_n)).
\end{align}

Similarly, a large CNF clause can be recursively decomposed use the following rule:

\begin{align}
(b_0\lor\cdots\lor b_n)
&\rightarrow
(b_0\lor\cdots\lor b_{\lfloor n/2\rfloor}\lor c)
\nonumber\\
&
\land
(b_{\lfloor n/2\rfloor+1}\lor\cdots\lor b_n\lor\overline c).
\end{align}

Each rewriting step approximately halves the maximum clause size, yielding lower-order Hamiltonian terms and smaller multi-controlled Toffoli gates.

\subsubsection{Dependency Analysis}

\begin{figure}[t!]
    \includegraphics[width=\linewidth]{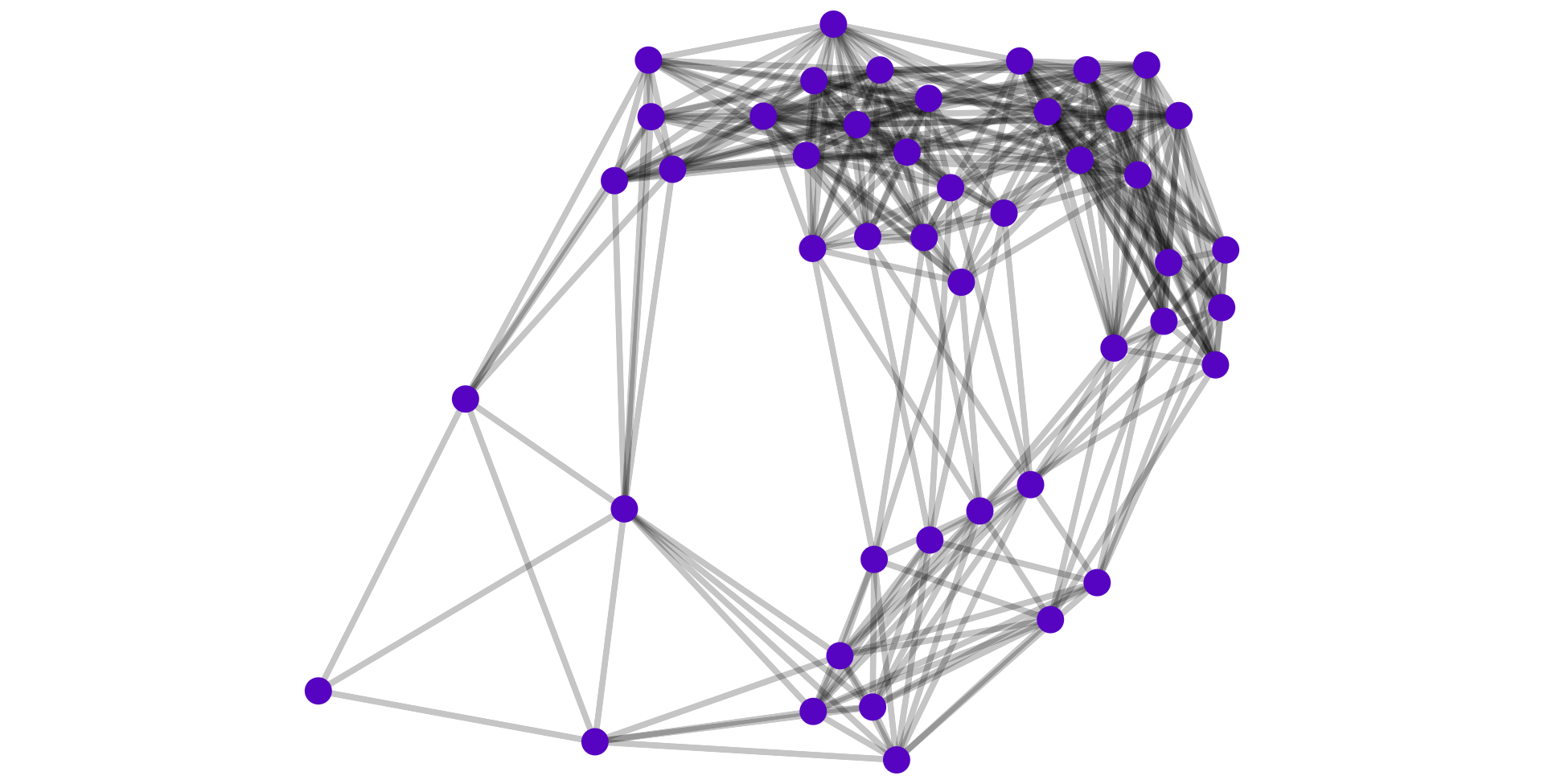}
    \caption {Connectivity of CNF clauses in a DIMACS miter circuit between a correct 1-bit full-adder circuit and one where one XOR gate has been replaced by XNOR.}
    \label{fig:FA_connectivity}
\end{figure}

Following rewriting, SAQC constructs a clause dependency graph
\[
G=(V,E),
\]
where each vertex represents a clause and an edge connects two clauses sharing at least one literal. Clauses without shared literals are independent and may therefore be executed in parallel. An example for a full-adder miter circuit is shown in \autoref{fig:FA_connectivity}.

\subsubsection{Clause Scheduling}\label{sec:clause-schedule}

The dependency graph is used to schedule independent clauses consecutively, reducing the QAOA ansatz depth. Since optimal scheduling is equivalent to the maximum independent set problem, SAQC employs a greedy heuristic with parameter $k$.

At each iteration, the algorithm selects the $k$ minimum-degree vertices, constructs their complement graph, and applies CLIPPER+~\cite{fathian2024clipper+} to compute a maximum clique, corresponding to the largest set of mutually independent clauses. These clauses are scheduled together, removed from the graph, and the procedure is repeated until all clauses have been processed. The complete algorithm is summarized in Algorithm~\autoref{alg:clause_schedule}.

\begin{figure}[t!]
    \begin{subfigure}{0.44\linewidth}
        \includegraphics[scale=0.8]{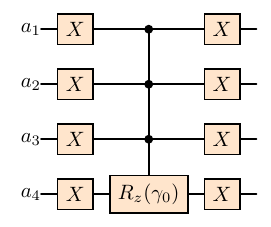}
        \caption{$a_1 \lor a_2 \lor a_3 \lor a_4$}
        \label{fig:decomp_or4}
    \end{subfigure}
    \begin{subfigure}{0.52\linewidth}
        \includegraphics[scale=0.8]{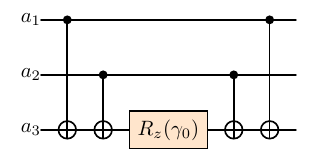}
        \caption{$a_1 \oplus a_2 \oplus a_3$}
        \label{fig:decomp_xor30}
    \end{subfigure}
    
    \vspace*{1.5em}
    \begin{subfigure}{0.42\linewidth}
        \includegraphics[scale=0.8]{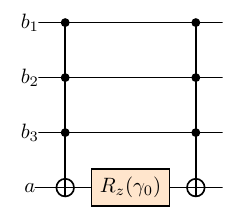}
        \caption{$a \oplus (b_1 \land b_2 \land b_3)$}
        \label{fig:decomp_xor13}
    \end{subfigure}
    \begin{subfigure}{0.6\linewidth}
        \includegraphics[scale=0.8]{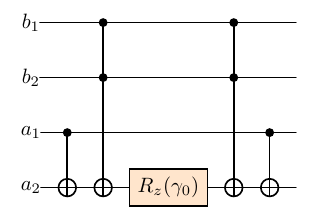}
        \caption{$a_1 \oplus a_2 \oplus (b_1 \land b_2)$}
        \label{fig:decomp_xor22}
    \end{subfigure}
    \caption{Direct dcomposition of different CNF and eCNF clauses into quantum gates.}
\end{figure}

\begin{algorithm}
\caption{Scheduling Algorithm}\label{alg:clause_schedule}
\begin{algorithmic}
\STATE $\text{\textbf{Input:} Clauses }V, \text{Clause-literals }L: L(v) \subseteq \text{Literals}, k$
\STATE $a(v_i, v_j) \Leftrightarrow L(v_i) \cap L(v_j) \neq \varnothing$ \\
\STATE $deg(v_i) \gets |\{ v_j \in V \mid  a(v_i, v_j)\}|$ \\
\WHILE{$|V| > 0$}
   \STATE Select $S \subseteq  V$ such that $|S| = k$ and $deg(v_i)$ is minimal \\
   \STATE $M \gets \max~\text{clique}(S)$ 
   \STATE Schedule $M$
   \STATE $V = V \setminus M$
\ENDWHILE
\end{algorithmic}
\end{algorithm}

\begin{figure*}[htbp!]
    \centering

    \begin{subfigure}{0.182\linewidth}
        \includegraphics[scale=0.8]{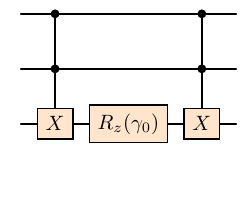}\vspace*{0.15em}
        \caption{}
        \label{fig:xor_base}
    \end{subfigure}
    \hspace{-0.3cm}
    \begin{subfigure}{0.24\linewidth}
        \includegraphics[scale=0.8]{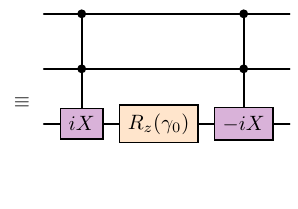}\vspace*{0.15em}
        \caption{}
        \label{fig:xor_rel}
    \end{subfigure}
    \hspace{-0.5cm}
    \begin{subfigure}{0.42\linewidth}
        \includegraphics[scale=0.8]{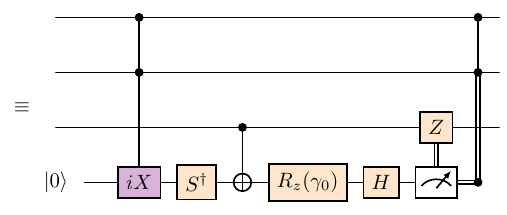}
        \caption{}
        \label{fig:xor_anc}
    \end{subfigure}
    \caption{Realizations of the $a \oplus (b\land c)$ rule using $(a)$ Toffoli Gates, $(b)$ Relative Phase Toffoli Gates and $(c)$ one ancilla with dynamic uncomputation.}
    \label{fig:xor}
    \vspace*{-1.5em}
\end{figure*}

\subsection{Lowering the SIR}

After SAT-aware optimization, the SIR is lowered through two complementary paths: (i) penalty Hamiltonian generation and (ii) direct clause-to-ansatz generation.

\subsubsection{Penalty Hamiltonian Generation}

Each SIR clause is lowered to a penalty Hamiltonian according to its logical operator. OR clauses are translated into projector Hamiltonians that penalize unsatisfied assignments, whereas XOR and XOR-of-product clauses are mapped using dedicated parity-preserving constructions. The corresponding lowering rules are

\begin{align}
(x_1\vee\cdots\vee x_n) &\rightarrow \prod_{i=1}^{n}\frac{I+Z_i}{2},\label{eq:nor}\\
x_1\oplus(x_2\land\cdots\land x_n) &\rightarrow I - \left[\frac{I-Z_1}{2}+\prod_{i=2}^{n-1}\frac{(I-Z_i)}{2}\right.\nonumber\\
& \qquad\qquad\left.- 2\prod_{i=1}^{n}\frac{(I-Z_i)}{2}\right], \label{eq:xor-and}\\
x_1\oplus\cdots\oplus x_n &\rightarrow \frac{1}{2}\left(I+\prod_{i=1}^{n}Z_i\right)\label{eq:nxor}.
\end{align}

For negated literals, the corresponding projector is obtained by interchanging
$\frac{I+Z_i}{2}$ and $\frac{I-Z_i}{2}$.
For XOR clauses, each complemented literal changes the parity of the constraint; consequently, the sign of the Pauli product in~\autoref{eq:nxor} is inverted whenever the number of complemented literals is odd. 

Since clause rewriting has already decomposed large Boolean expressions, the generated Hamiltonian contains lower-order interactions, resulting in a more compact representation and shallower circuits after Hamiltonian synthesis.

\subsubsection{Direct Clause-to-Ansatz Generation}

Besides penalty-Hamiltonian generation, SAQC directly synthesizes the QAOA cost layer from the optimized SIR. Instead of constructing
$U_C(\gamma)=e^{-i\gamma H_C}$
from the clause Hamiltonians, each clause is mapped directly to a parameterized gate template.

For CNF clauses, an OR constraint is realized using a multi-controlled $R_z$ gate. By De Morgan's law, positive literals are first negated, followed by the controlled rotation and restoration of the original basis, as illustrated in~\autoref{fig:decomp_or4}.

For eCNF clauses, XOR expressions are implemented using CX chains that accumulate the parity onto a target qubit before applying an $R_z$ rotation (e.g., see~\autoref{fig:decomp_xor30}). XOR-of-product expressions additionally employ multi-controlled Toffoli gates to compute temporary conjunctions prior to the phase rotation, as illustrated in
Figs.~\ref{fig:decomp_xor13}--\ref{fig:decomp_xor22}.

Operating directly on the SIR avoids constructing large intermediate Hamiltonian expressions while naturally supporting the SAT-aware optimizations introduced in the previous subsection.

\subsection{Ansatz Optimizations}

The clause-to-gate translation produces a correct QAOA ansatz but may introduce large multi-controlled Toffoli gates, resulting in high decomposition cost. To improve the generated circuits, SAQC applies three complementary optimizations: relative-phase decomposition, ancilla-assisted decomposition, and dynamic uncomputation.

\subsubsection{Relative-phase Decomposition}

The XOR-based construction in \autoref{fig:xor}(a) realizes an eCNF clause of the form
\[
a_1\oplus\cdots\oplus a_n\oplus(b_1\land\cdots\land b_m)
\]
using an $R_z$ gate and an $m$-controlled Toffoli gate. Since decomposing multi-controlled Toffoli gates into the Clifford+$T$ gate set is expensive, SAQC replaces them with relative-phase variants. As each conjunction is computed and subsequently uncomputed, the relative phases introduced during computation are cancelled by the inverse relative-phase Toffoli during uncomputation. This preserves the logical functionality while reducing the number of CX and T gates, as illustrated in \autoref{fig:xor}(b).

\subsubsection{Ancilla-assisted Decomposition}

Large conjunctions are further optimized using clean ancilla qubits. Instead of directly realizing an $m$-controlled Toffoli gate, the conjunction is first computed onto an ancilla, followed by the phase rotation and subsequent uncomputation. This replaces large multi-controlled gates with smaller Toffoli gates, reducing decomposition cost.

The effectiveness of this optimization depends on the available ancilla budget. Reusing a single ancilla serializes clause execution, whereas sufficient ancillas enable independent clauses identified during scheduling (\autoref{sec:clause-schedule}) to execute in parallel.

\subsubsection{Dynamic Uncomputation}

The ancilla-assisted construction is further optimized using dynamic uncomputation. Instead of reversing the complete compute network, the ancilla is measured in the Hadamard basis and removed using classically controlled $CZ$ and $Z$ corrections, as shown in \autoref{fig:xor}(c). This replaces much of the inverse circuit with a few native operations, reducing both gate count and circuit depth. Combined with relative-phase decomposition and ancilla-assisted synthesis, this optimization produces significantly more resource-efficient QAOA ansätze.
\section{Experimental Evaluation}

\begin{table*}[tbh]
\begin{center}
\caption{Differences in depth between the Qiskit ansatz and our approach. 
    \label{tbl:depth-comparison-1}}
    \begin{tabular}{l|rrr|rrrrr|rrrrr}
    &&&Largest&\multicolumn{5}{c|}{0 Ancilla}
    &\multicolumn{5}{c}{15 Ancilla}
    \\
    Benchmark & Clauses & Variables & Clause &
    Qiskit & \emph{time} & Us & \emph{time} &Diff\% &
    Qiskit & \emph{time} & Us & \emph{time} &Diff\% \\
    \hline
s1\_mul\_4\_4 & 421 & 428 & 9 & 2071 &  0.94s  & 236 & 0.19s & 88.60\% & 177 &  0.31s  & 144 & 0.17s & 18.64\% \\
s1\_mul\_5\_5 & 757 & 766 & 11& 10287 &  1.78s  & 219 & 0.42s & 97.87\% & 223 &  0.63s  & 176 & 0.30s & 21.08\% \\
s1\_mul\_6\_6 & 1164 & 1175 & 13 & 49287 &  6.30s  & 303 & 1.09s & 99.39\% & 278 &  1.28s  & 206 & 0.91s & 25.90\% \\
s1\_mul\_7\_7 & 1683 & 1696 & 15 & 229856 &  57.90s  & 316 & 2.22s & 99.86\% & 336 &  2.64s  & 227 & 1.87s & 32.44\% \\
s1\_mul\_8\_8 & 2276 & 2291 & 17 & \emph{N/A} &    & 393 & 2.89s &  & 399 &  4.59s  & 274 & 2.40s & 31.33\% \\
s1\_mul\_9\_9 & 2947 & 2964 & 19 & \emph{N/A} &    & 315 & 6.03s &  & 459 &  7.01s  & 317 & 5.35s & 30.94\% \\
    s1\_mul\_10\_10 & 3756 & 3775 & 21 & \emph{N/A} &    & 432 & 5.49s &  & 510 &  11.06s  & 332 & 4.51s & 34.90\% \\
s1\_mul\_11\_11 & 4582 & 4603 & 23 & \emph{N/A} &    & 373 & 7.52s &  & 561 &  16.48s  & 360 & 6.54s & 35.83\% \\
cl\_4 & 143 & 150 & 6& 173 &  0.67s  & 113 & 0.09s & 34.68\% & 74 &  0.07s  & 82 & 0.07s & -10.81\% \\
cl\_8 & 327 & 342 & 10 & 4638 &  0.93s  & 133 & 0.14s & 97.13\% & 79 &  0.16s  & 82 & 0.08s & -3.80\% \\
cl\_16 & 695 & 726 & 18 & \emph{N/A} &    & 162 & 0.41s &  & 101 &  0.55s  & 86 & 0.29s & 14.85\% \\
    \end{tabular}\\
    \vspace*{0.4em}
    {\footnotesize  Clauses marked with \emph{N/A} exceeded the time budget of 20 minutes during construction.}
    \end{center}
\end{table*}

\begin{table}[tbh]
\begin{center}
\caption{Comparison of two-qubit gate counts.\label{tbl:cx-comparison}}
    \hspace*{-0.8em}\begin{tabular}{l|rrr|rrr}
    &\multicolumn{3}{c|}{0 Ancilla}
    &\multicolumn{3}{c}{15 Ancilla}
    \\
    Benchmark &
    Qiskit & Us & Diff\% &
    Qiskit & Us & Diff\% \\
    \hline

s1\_mul\_4\_4 &  4088 & 1778 & 56.51\% & 2222 & 1772 & 20.25\% \\
s1\_mul\_5\_5 & 13622 & 3166 & 76.76\% & 3964 & 3160 & 20.28\% \\
s1\_mul\_6\_6 & 53094 & 4848 & 90.87\% & 6076 & 4842 & 20.31\% \\
s1\_mul\_7\_7 & 229830 & 6978 & 96.96\% & 8748 & 6972 & 20.30\% \\
s1\_mul\_8\_8 & \emph{N/A} & 9424 &  & 11820 & 9418 & 20.32\% \\
s1\_mul\_9\_9 & \emph{N/A} & 12186 &   & 15294 & 12180 & 20.36\% \\
s1\_mul\_10\_10 & \emph{N/A} & 15536 &  & 19518 & 15530 & 20.43\% \\
s1\_mul\_11\_11 & \emph{N/A} & 18900 &  & 23746 & 18894 & 20.43\% \\
\hline
cl\_4 & 812 & 574 & 29.31\% & 698 & 568 & 18.62\% \\
cl\_8 &  5892 & 1310 & 77.77\%  & 1602 & 1304 & 18.60\% \\
cl\_16 &  \emph{N/A} & 2798 &   & 3434 & 2792 & 18.70\% \\

    \end{tabular}\\\vspace*{0.4em}
    {\footnotesize  Clauses marked with \emph{N/A} exceeded the time budget of 20 minutes during construction.}
    \end{center}
\end{table}



Experiments were performed on eCNF SAT benchmarks from~\cite{10.1145/3729229} using an Intel Core Ultra~7~255U CPU with 64\,GB RAM. Randomized experiments were averaged over five runs. SAQC is compared against the Hamiltonian-based Qiskit workflow, with all circuits transpiled to the native gate set $\{CX,R_Z,S,T,H,X\}$ (optimization level~3). Unless otherwise stated, the clause scheduling parameter is $k=1200$.

\subsection{Comparison with Qiskit}

Table~\ref{tbl:depth-comparison-1} compares the circuit depth and compilation time of SAQC and the Hamiltonian-based Qiskit workflow using both 0 and 15 additional ancilla qubits. Without auxiliary qubits, SAQC consistently produces substantially shallower circuits, achieving depth reductions of up to $99.87\%$. When clause rewriting is enabled using 15 ancillas, the Hamiltonian-based workflow benefits from smaller Hamiltonians, reducing the performance gap for small benchmarks. Nevertheless, SAQC continues to outperform Qiskit on larger instances, achieving up to $35.83\%$ lower circuit depth.

In addition to circuit depth, SAQC significantly reduces compilation time by avoiding explicit Hamiltonian construction. For example, on the $s1\_mul\_11\_11$ benchmark, the total compilation time decreases from $16.57$\,s for the Qiskit workflow to $6.54$\,s using SAQC. Clause rewriting itself contributes negligible overhead, requiring less than $0.2$\,s across all benchmarks.

Table~\ref{tbl:cx-comparison} reports the corresponding two-qubit gate counts. SAQC consistently requires fewer CX gates, achieving reductions of up to $96.96\%$ without ancillas and approximately $20\%$ even when clause rewriting is enabled.

\begin{figure}
    \includegraphics[trim={0cm 0.2cm 0cm 0.38cm}, clip, width=0.95\linewidth]{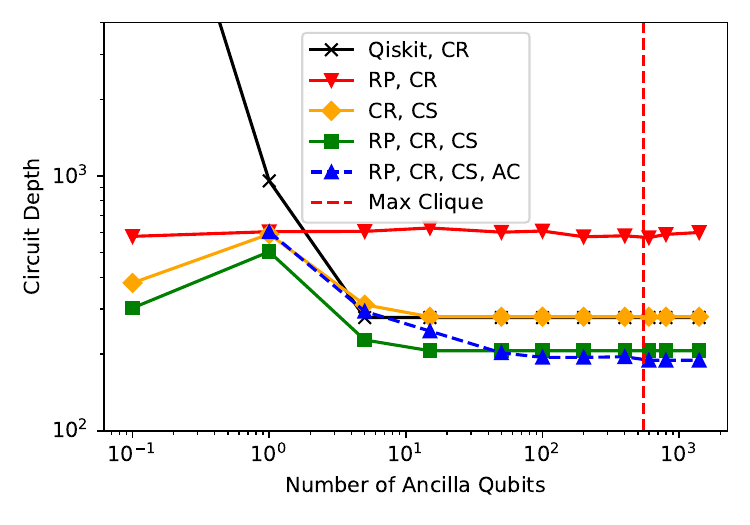}
    \caption{Comparison of Circuit Depth with ancilla increasing towards max\_clique. 
    \emph{CR}:~\emph{Clause Rewriting},
    \emph{CS}:~\emph{Clause Scheduling},
    \emph{RP}:~\emph{Relative-Phase Toffoli Construction},
    \emph{AC}:~\emph{Ancilla-Based Construction},
    \label{fig:adaptive_rewriting_depth}}
\end{figure}

\subsection{Impact of Individual Optimizations}

To evaluate the contribution of each optimization, Fig.~\ref{fig:adaptive_rewriting_depth} reports the circuit depth of the $6$-bit multiplier benchmark while progressively enabling clause rewriting, clause scheduling, relative-phase decomposition, and ancilla-assisted decomposition.

Clause scheduling provides the largest reduction in circuit depth by exposing parallelism between independent clauses. Relative-phase Toffoli decomposition further decreases the depth by reducing the cost of multi-controlled gates. Ancilla-assisted decomposition exhibits a more nuanced behavior: when only a few ancillas are available, the additional serialization outweighs its benefits, whereas near the maximum independent set size it achieves the lowest circuit depth. This trend is consistent across all evaluated benchmarks.

A small depth increase is observed for configurations with only a few auxiliary qubits. Although the precise cause remains under investigation, we attribute this behavior to heuristics employed by the Qiskit transpiler when decomposing multiple smaller Toffoli gates.

\section{Conclusion}
This paper presented SAQC, a SAT-aware compilation framework for QAOA that introduces a SAT Intermediate Representation (SIR) to preserve Boolean semantics prior to quantum lowering.

Building on the SIR, SAQC provides a unified flow for penalty Hamiltonian generation and direct clause-to-ansatz synthesis from CNF and eCNF formulations. Combined with SAT-aware optimization passes and ansatz-level optimizations, the framework generates more resource-efficient QAOA circuits while remaining compatible with existing quantum compilation frameworks.

Experimental results demonstrate significant reductions in circuit depth, two-qubit gate count, and compilation time compared with conventional Hamiltonian-based workflows. Future work will extend SAQC to richer Boolean constraints and hardware-aware compilation for emerging quantum architectures.

\section{Acknowledgments}
\thanks{This work was partly funded by the Federal Ministry of Research, Technology and Space (BMFTR) through the EASEPROFIT project (grant no. 16KIS2127) and by the German Research Foundation (DFG) through the CONAD-QC project (grant no. 559888852). The research is conducted within the scope of the DFG Priority Programme 2514 (SPP 2514).}

\bibliographystyle{IEEEtran}
\bibliography{bib/QAOA,bib/SAT}

\end{document}